\documentclass[conference]{IEEEtran}
\IEEEoverridecommandlockouts
\usepackage{amsmath,amssymb,amsfonts}
\usepackage{graphicx}
\usepackage{textcomp}
\usepackage{xcolor}
\usepackage{xcolor,soul,framed} 
\usepackage{fancyvrb}
\usepackage{relsize}
\usepackage{listings}

\usepackage{booktabs}
\colorlet{shadecolor}{yellow}

\usepackage{moreverb}
\usepackage[hidelinks]{hyperref}
\usepackage[numbers,sort&compress]{natbib}
\usepackage{url}

\usepackage{algorithm}
\usepackage{algpseudocode}

\def\BibTeX{{\rm B\kern-.05em{\sc i\kern-.025em b}\kern-.08em
    T\kern-.1667em\lower.7ex\hbox{E}\kern-.125emX}}

\makeatletter
\newcommand{\linebreakand}{%
  \end{@IEEEauthorhalign}
  \hfill\mbox{}\par
  \mbox{}\hfill\begin{@IEEEauthorhalign}
}
\makeatother
\begin{document}
\title{
Benefits of Dynamic Thermal Rating for Distribution Transformers Under Accelerated Load Growth\\

\thanks{
This work was authored by the National Laboratory of the Rockies for the U.S. Department of Energy (DOE), operated under Contract No. DE-AC36-08GO28308. Funding provided by Department of Energy office of Electricity. The views expressed in the article do not necessarily represent the views of the DOE or the U.S. Government. The U.S. Government retains and the publisher, by accepting the article for publication, acknowledges that the U.S. Government retains a nonexclusive, paid-up, irrevocable, worldwide license to publish or reproduce the published form of this work, or allow others to do so, for U.S. Government purposes.

}
}

\author{

\IEEEauthorblockN{Wenbo Wang, Killian McKenna}
\IEEEauthorblockA{\textit{Grid Planning and Analysis Center} \\
\textit{National Laboratory of the Rockies }\\
Golden, CO, USA \\
\{wenbo.wang, killian.mckenna\}@nlr.gov}

\and

\IEEEauthorblockN{Francisco de Leon}
\IEEEauthorblockA{\textit{ECE Department} \\
\textit{New York University}\\
Brooklyn, NY, USA \\
fdeleon@nyu.edu}



\and

\IEEEauthorblockN{Haowei Lu}
\IEEEauthorblockA{\textit{Technology Engineering} \\
\textit{Orange \& Rockland Utilities}\\
Spring Valley, NY, USA \\
luh1@oru.com}

\and

\IEEEauthorblockN{Xin Fang}
\IEEEauthorblockA{\textit{EE Department} \\
\textit{University of South Carolina}\\
Columbia, SC, USA \\
fangxin@sc.edu }

}

\maketitle

\begin{abstract}

Dynamic Thermal Rating (DTR) has been widely applied to transmission lines for congestion management, but its potential for distribution transformers is less explored. By adjusting equipment ratings based on real ambient and operating conditions, DTR allows transformers to operate above conservative static nameplate values without compromising asset health. This paper evaluates how DTR can serve as a non-wires alternative to traditional upgrades in distribution systems facing accelerated load growth. Using an example feeder from the SMART-DS dataset, we compare conventional transformer replacement strategies against DTR-enabled operation with ambient-adjusted ratings. The results demonstrate that DTR can defer a significant number of transformer upgrades and reduce annual upgrade costs. More broadly, DTR is most effective in regions with pronounced weather diversity, peak demand occurring during cooler periods, or sustained load growth that would otherwise trigger costly reinforcements. These findings show that DTR is a practical and cost-effective option for utilities, provides both operational flexibility and investment deferral opportunities, and offers a foundation for developing distribution-level DTR programs.

\end{abstract}


\section{Introduction}


The transfer capacity of power distribution equipment is traditionally determined by calculating the maximum load it can handle without overheating based on the most conservative ambient weather conditions (e.g., full sun in the afternoon, low wind speeds and historic maximum ambient temperature in the region). But dynamic thermal rating (DTR) offers a more adaptive approach by assessing equipment capacity considering realistic weather conditions over different time scales (e.g., hourly, weekly, seasonally) \cite{dlr_congress} without compromising the safety of operation and without a negative impact on the life expectancy of the equipment. As a result, DTR provide utilities with real-time and location-specific asset ratings rather than relying on static and conservative estimates. The use of DTR could address emerging challenges in distribution grids, particularly those arising from accelerated load growth driven by residential electrification from heat pumps and electric vehicles (EVs) \cite{Mai_NREL_EFS}, as well as increasing cooling and heating demands. For instance, heat pumps typically represent a higher load than most common household appliances, especially when running at full capacity, potentially exceeding the rated capacity of the upstream distribution transformers. In addition, prior studies have shown that distribution transformer capacity can become the bottleneck for supporting EV charging \cite{EV_impact_2003}. While upgrading network assets is essential for large load growth, utilities can first leverage DTR as a faster, cost-effective solution to manage moderate load increases, delaying significant infrastructure investments without accelerating equipment aging. Reference \cite{IET_HC_EV_charging} evaluates the EV hosting capacity (as the maximum number of EV chargers that can be safely connected) in distribution networks considering static ratings and the DTRs of the substation transformer and primary overhead lines.

\begin{figure}[t!]
    \centering
    \includegraphics[scale=0.43]{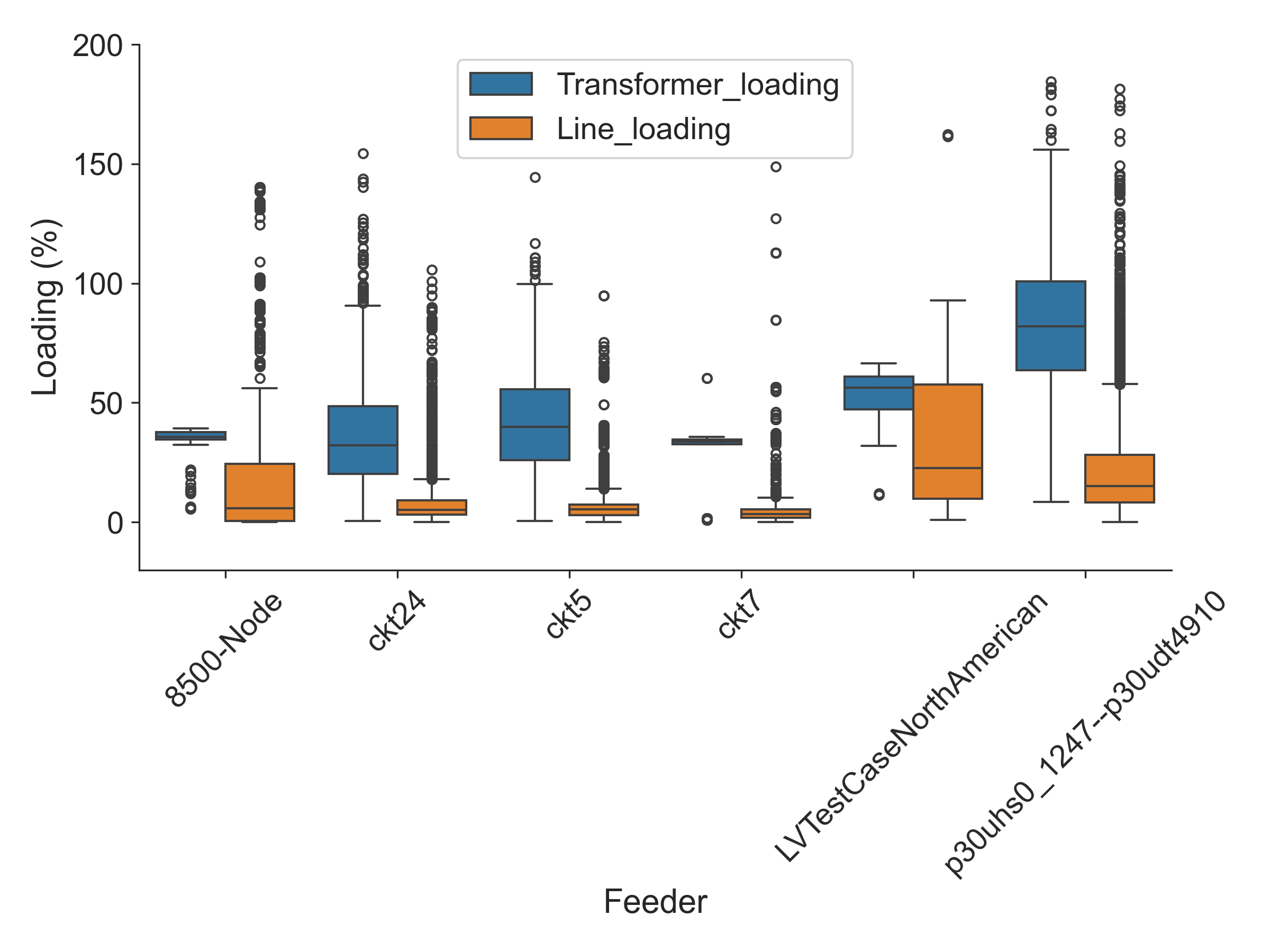}
    \caption{Boxplot showing the percentage loading of distribution transformers and lines across various test feeders from the IEEE, EPRI, and SMART-DS databases. }
    \label{fig:boxplot_loading_all}
\end{figure}

While load growth affects all classes of transformers, this paper focuses on distribution service transformers, which are traditionally less monitored yet, deployed in large numbers. These assets are increasingly stressed by localized electrification and customer growth, making them a critical but often overlooked bottleneck. In fact, prior studies have shown that distribution transformer capacity is often the key limiting factor for supporting new loads \cite{EV_impact_limiting_factor}, \cite{EVHC_Microgrids}. This is consistent with our own snapshot power flow analysis using IEEE, EPRI, and SMART-DS feeders that include secondary models. As illustrated in Fig.~\ref{fig:boxplot_loading_all}, distribution transformers operate closer to their ratings compared with both primary and secondary lines. The boxplot shows that transformer loading is consistently higher across all feeders, confirming that service transformers frequently represent the binding constraint in U.S. distribution networks.


The extent of DTR benefits is directly related to how effectively the additional capacity is applied to system needs. For example, at the transmission level, dynamic line rating has been used to relieve congestion and reduce operating costs~ \cite{dlr_congress}. At the distribution level, the benefit can be measured against the costly alternative of transformer upgrades. Although existing research has highlighted the potential of DTR, few studies directly compare it with traditional distribution transformer replacements. The contribution of this paper is a quantified evaluation of DTR at the distribution level using a representative SMART-DS feeder, with results that justify its consideration as a practical and cost-effective option in utility planning.

\section{DTR Models for Distribution Transformers}


\begin{figure}[t!]
    \centering
    \includegraphics[scale=0.5]{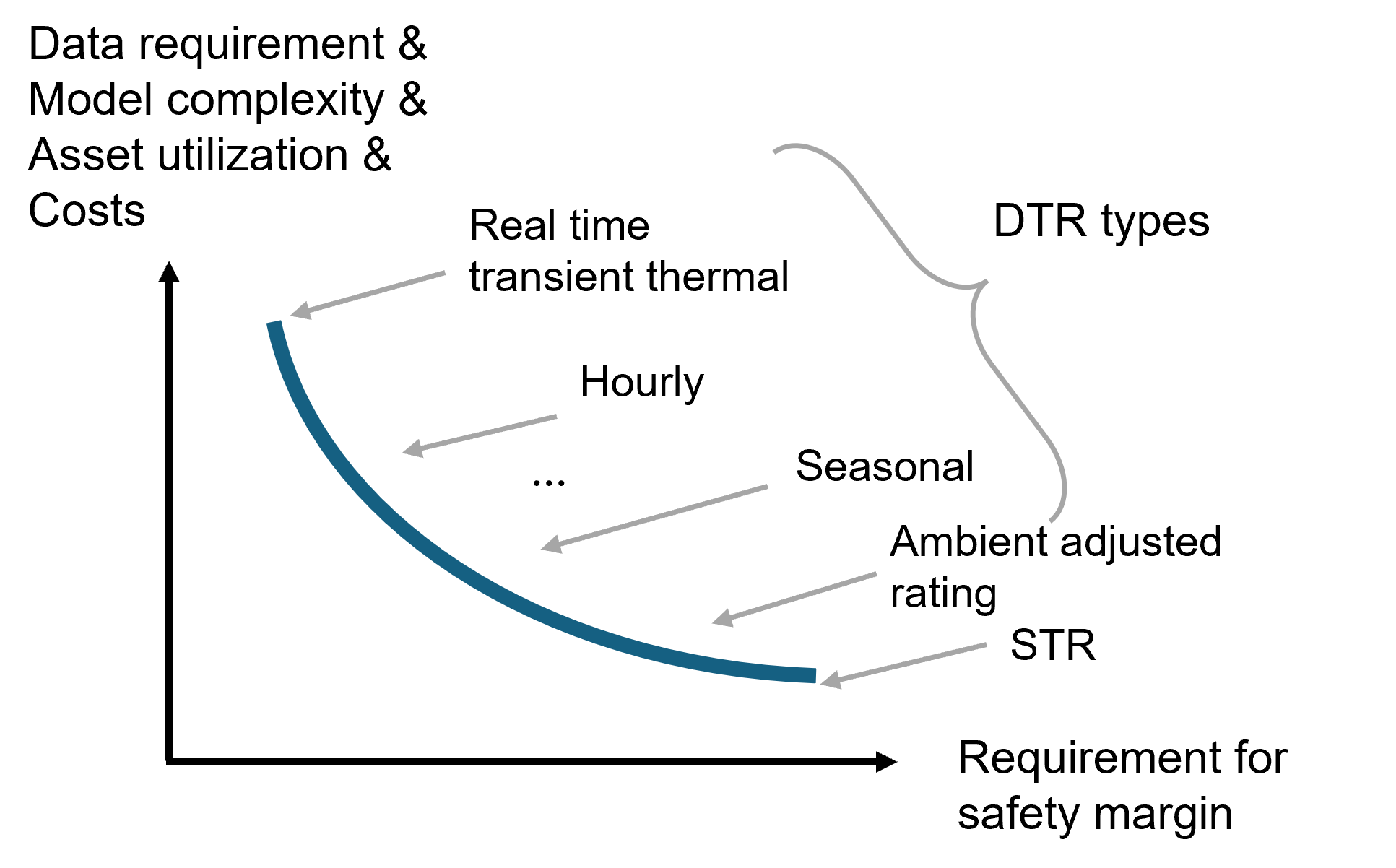}
    \caption{Comparison of DTR types highlighting the trade-off between data requirement, model complexity, and safety margins requirement.}
    \label{fig:DTR_types}
\end{figure}
\begin{figure}[t!]
    \centering
    \includegraphics[scale=0.4]{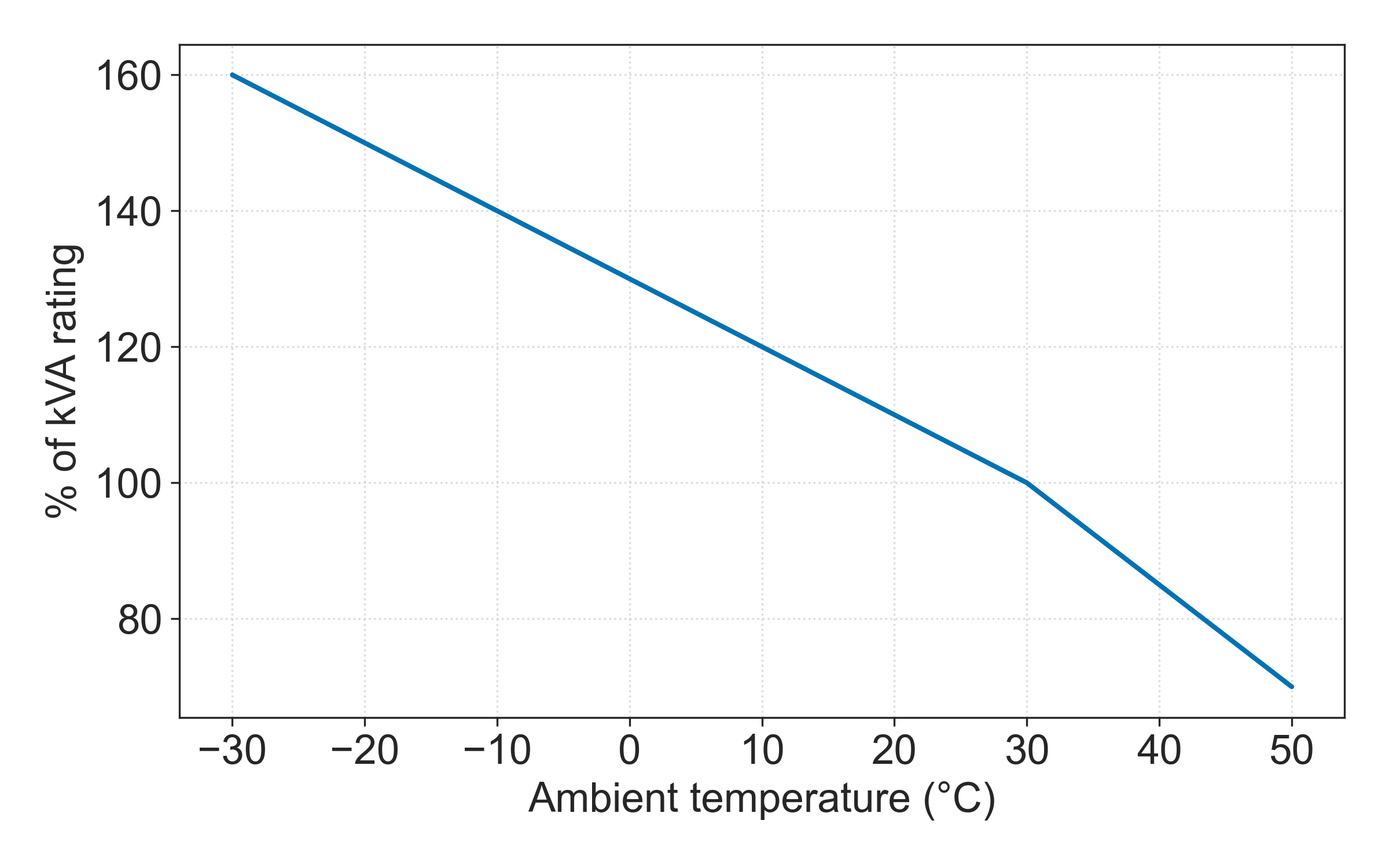}
    \caption{Ambient adjusted rating estimates based on ambient temperature on the basis of 25$^\circ$C}
    \label{fig:AAR_model_IEEE}
\end{figure}


DTR encompasses a spectrum of approaches that vary in timescale, model complexity, and data requirements. At one end, simple methods such as ambient-adjusted rating (AAR) apply temperature-based adjustment factors \cite{ieee_std_5791}, \cite{ieee_std_5796}. This approach is easy to implement and requires minimal data, but still needs some safety margins. At the other end, real-time transient thermal models and hourly models continuously calculate thermal states based on detailed weather inputs and transformer parameters. These approaches maximize asset utilization but require extensive sensing, communications, and integration, as well as higher implementation costs. Between these extremes are intermediate options such as weekly and monthly rating models, which provide a balance between accuracy and complexity. Fig.~\ref{fig:DTR_types} illustrates this trade-off: as model complexity and data requirements increase, the safety margin decreases, enabling higher utilization of transformer capacity. In this study, we adopt the AAR method for oil-natural air-natural transformers \cite{ieee_std_5791}, as it represents the simplest and most practical form of DTR to implement, and can also serve as a transitional step toward more advanced real-time DTR methods.

\begin{figure}[t!]
    \centering
    \includegraphics[scale=0.38]{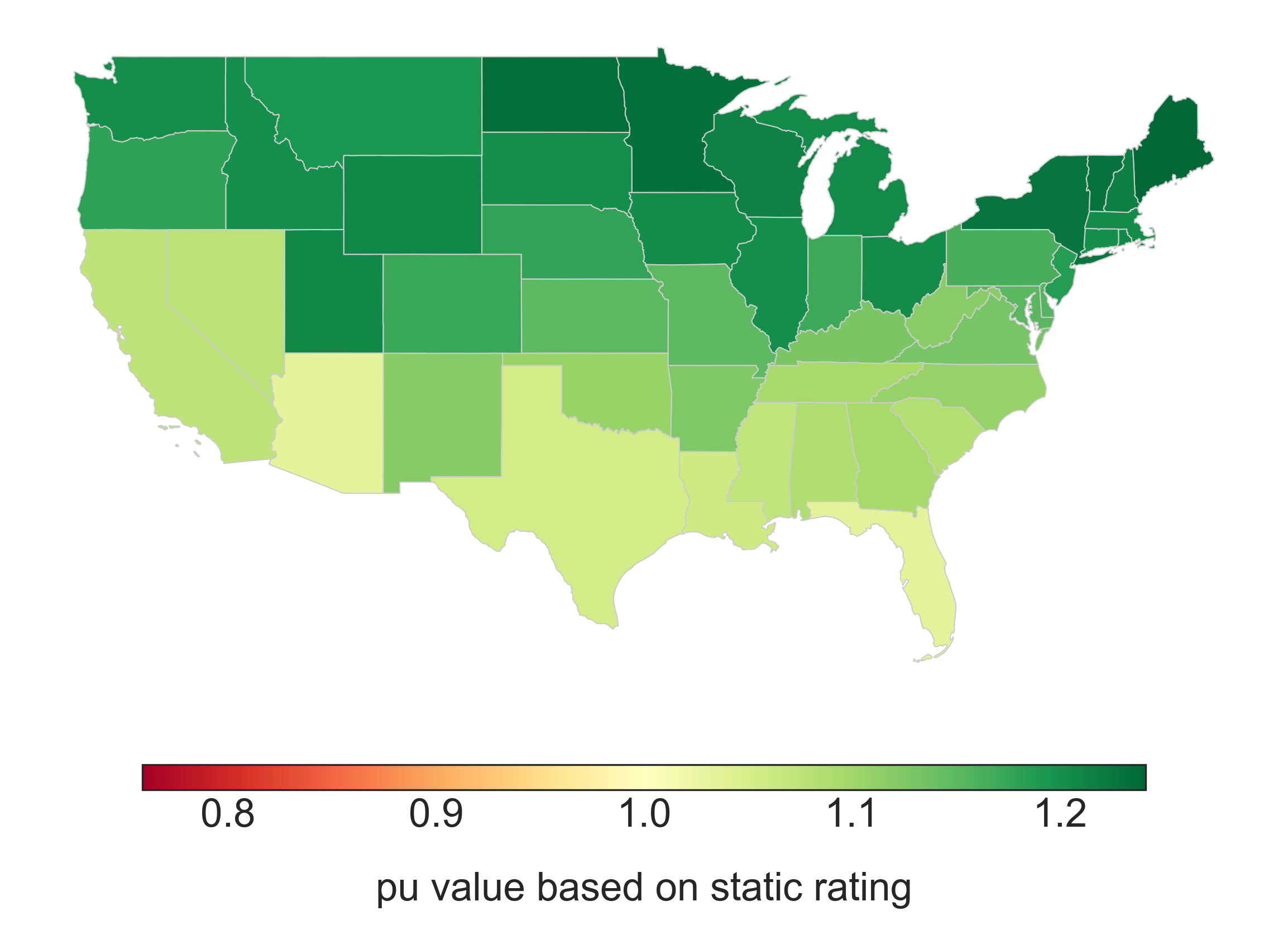}
    \caption{Heatmap showing ambient adjusted rating potential of distribution transformers in winter, with pu values on the basis of static ratings. Northern states such as Maine and Minnesota have a higher potential for DTR, i.e., 125 percent increase. Temperature data are from the Sup3rCC dataset for the year 2023, which was notably hot.} 
    \label{fig:Dtrans_DTR_potential_winter}
\end{figure}

AAR estimates the variation in rated kVA loading based on ambient temperature. In most cases, this leads to higher ratings than the static ratings. For instance, because static ratings assume high ambient temperatures, transformers can often be rated higher during winter when actual temperatures are lower. The model provided in \cite{ieee_std_5791}, illustrated in Fig. \ref{fig:AAR_model_IEEE}, shows the AAR estimates based on ambient temperature. This model also maintains approximately the same life expectancy as if the transformers were operated at the nameplate rating and standard ambient temperatures over the same period. As shown in the figure, for temperatures above 30$^\circ$C, the rating decreases by 1.5\% for each degree that the ambient temperature increases, and the rating increases 1\% for each degree that the temperature decreases. In this paper, we calculate AAR for each hour and select 5\% quantile over selected periods such as seasonal AAR. Using the National Renewable Energy Laboratory (NREL) Sup3rCC \cite{NREL_supercc} dataset, which provides high-resolution climate data for the contiguous United States, we calculate the AAR for the region. Fig. \ref{fig:Dtrans_DTR_potential_winter} presents the geographic color map illustrating the AAR potential during winter months in terms of per unit values based on static ratings. This is particularly advantageous for utilities experiencing winter peak demand, e.g., Holy Cross Energy's (a utility in Colorado) winter load is particularly sensitive to weather, as resort snowmaking requires specific cold weather conditions, and resort visitation fluctuates based on snow conditions \cite{holycross2023}. For other types of utilities, the benefits of DTR could extend beyond increased ratings, providing increased situational awareness of weather impacts on the grid. Although seasonal (summer/winter) ratings are widely used by many transmission operators and distribution utilities for power transformer assets \cite{eversource_planning_guide}, their application to distribution transformers remains less common \cite{seasonal_rating_canada}, \cite{aep2024}.
This study provides a justification of adopting more spatial-temporal granular ratings for distribution transformers or further deploying monitoring devices, as suggested in \cite{paloalto2018}.

\section{Case Study: Distribution Transformer Upgrade vs. DTR}

\subsection{Distribution Transformer Upgrade Cost}

The transformer upgrade cost includes both direct and indirect costs. The direct cost refers to the purchase of the transformer, while the indirect costs include transportation, installation, site preparation, old transformer removal, system integration, and other related expenses. Here, we assume that the purchase cost is based on the NREL distribution grid integration unit cost database \cite{NREL_unit_cost_database}. We also assume that the indirect costs are 100\% of the direct cost, a figure that matches a few other data sources, such as data from Eversource guide \cite{eversource}
and Southern California Edison \cite{sce}. 



\begin{figure}[t!]
    \centering
    \includegraphics[scale=0.4]{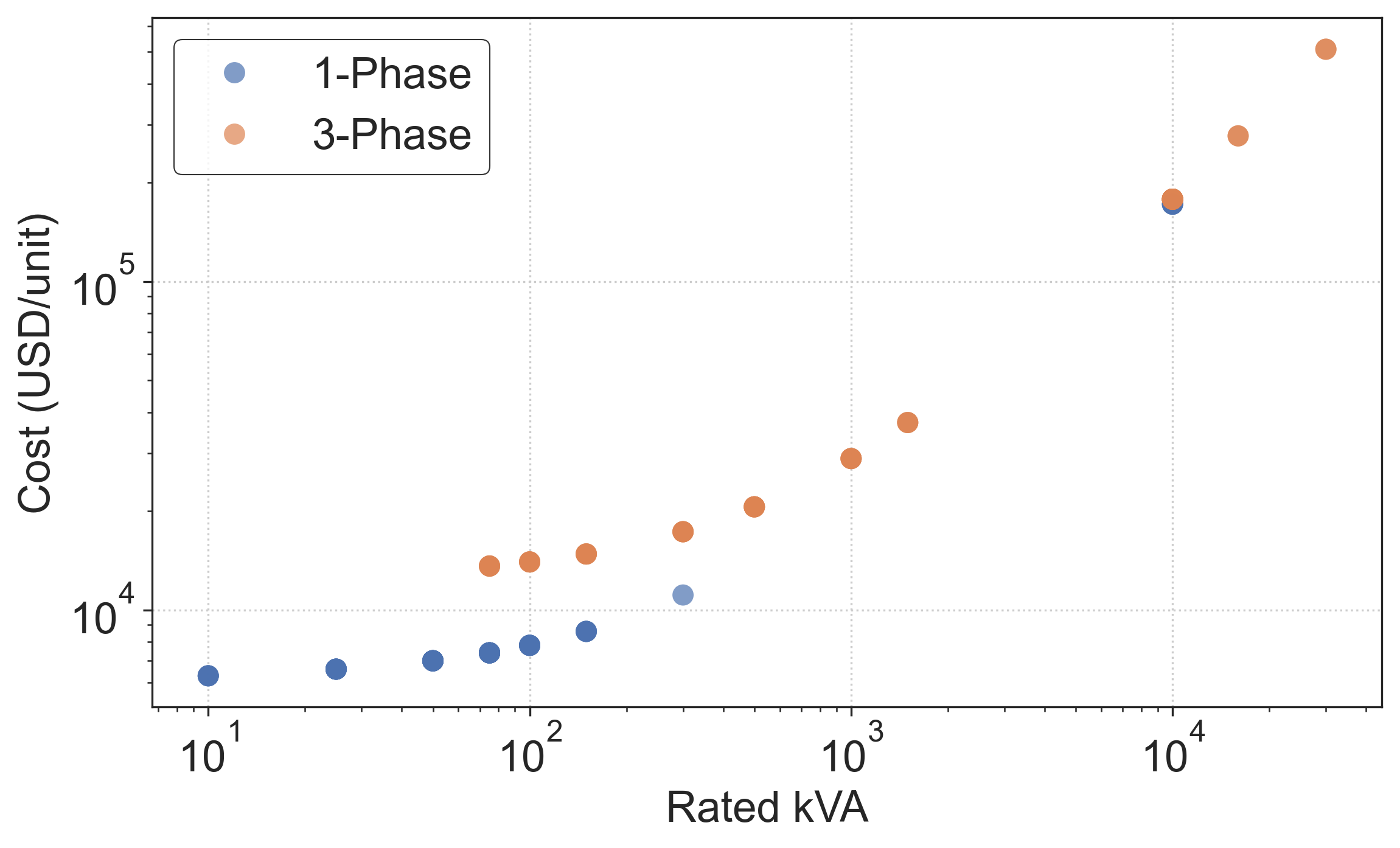}
    \caption{Purchase cost from NREL unit database}
    \label{fig:cost_vs_kva}
\end{figure}

Fig. \ref{fig:cost_vs_kva} shows the relationship between the rated kVA and the cost per unit in a log-log scale, allowing for the inclusion of very large kVA ratings. Because this cost database does not cover all the rated kVA transformers, linear regressions were performed for one-phase and three-phase transformers, respectively. The resulting functions are as follows:
\begin{equation}
\text{purchase\_cost\_1p} = 16.58 \times \text{rated\_kva} + 6156.19
\label{eq:1phase}
\end{equation}
\begin{equation}
\text{purchase\_cost\_3p} = 16.58 \times \text{rated\_kva} + 12324.16
\label{eq:3phase}
\end{equation}

Equations \eqref{eq:1phase} and \eqref{eq:3phase} indicate that the cost increases linearly with the rated kVA for both single-phase and three-phase transformers. The slope of the line is approximately the same for both, but the intercept is higher for three-phase transformers, reflecting a higher base cost. Both R-squared values from the linear regressions are close to 0.99. 


\subsection{Annual Utility Distribution Transformer Upgrade Modeling}


Under a load growth scenario of 5\% per year, uniformly applied to all loads, a transformer is assumed to be replaced once it peak load reaches 100\% of its rating. The upgraded transformer is sized such that its loading does not exceed 75\% of its kVA capacity. In practice, this means selecting the smallest available size from the standard list $(10, 15, 25, 50, 75, 100, 150, 167, 225, 250, 300, 500, ...)$ that satisfies the 75\% loading criterion. The pseudocode for the traditional upgrade procedure is presented in Algorithm \ref{alg:replacement}.

\begin{algorithm}[htbp]
\caption{Transformer Replacement Simulation} \label{alg:replacement}
\begin{algorithmic}[1] 
\State Initialize OpenDSS model and base loads
\For{year = 1 to 15}
    \State Apply load growth: $L \gets L_{\text{base}} \cdot (1 + r)^{(year - 1)}$
    \State Solve power flow
    \State Get transformer loading data
    \ForAll{transformers where loading $>$ threshold}
        \State Calculate required size for 75\% target loading
        \State Select standard transformer size
        \State Replace transformer in OpenDSS model
        \State Record replacement cost and details
    \EndFor
    \State Re-solve power flow with new transformer sizes
    \State Verify all overloads resolved
    \State Save yearly results
\EndFor
\end{algorithmic}
\end{algorithm}

An example feeder named $p30udt4910-p30uhs0\_1247x$ in the SMART-DS dataset is selected for demonstration\footnote{The dataset is publicly available at  \url{https://data.openei.org/submissions/2981}}. It is an urban residential area with a primary voltage of 12.47 kV, a total peak load of 12.5 MW, and a total capacity of 29.6 MW. The feeder includes 615 service transformers, mostly single-phase, with an average of 6 loads per transformer. 

\begin{figure}[t!]
    \centering
    \includegraphics[scale=0.35]{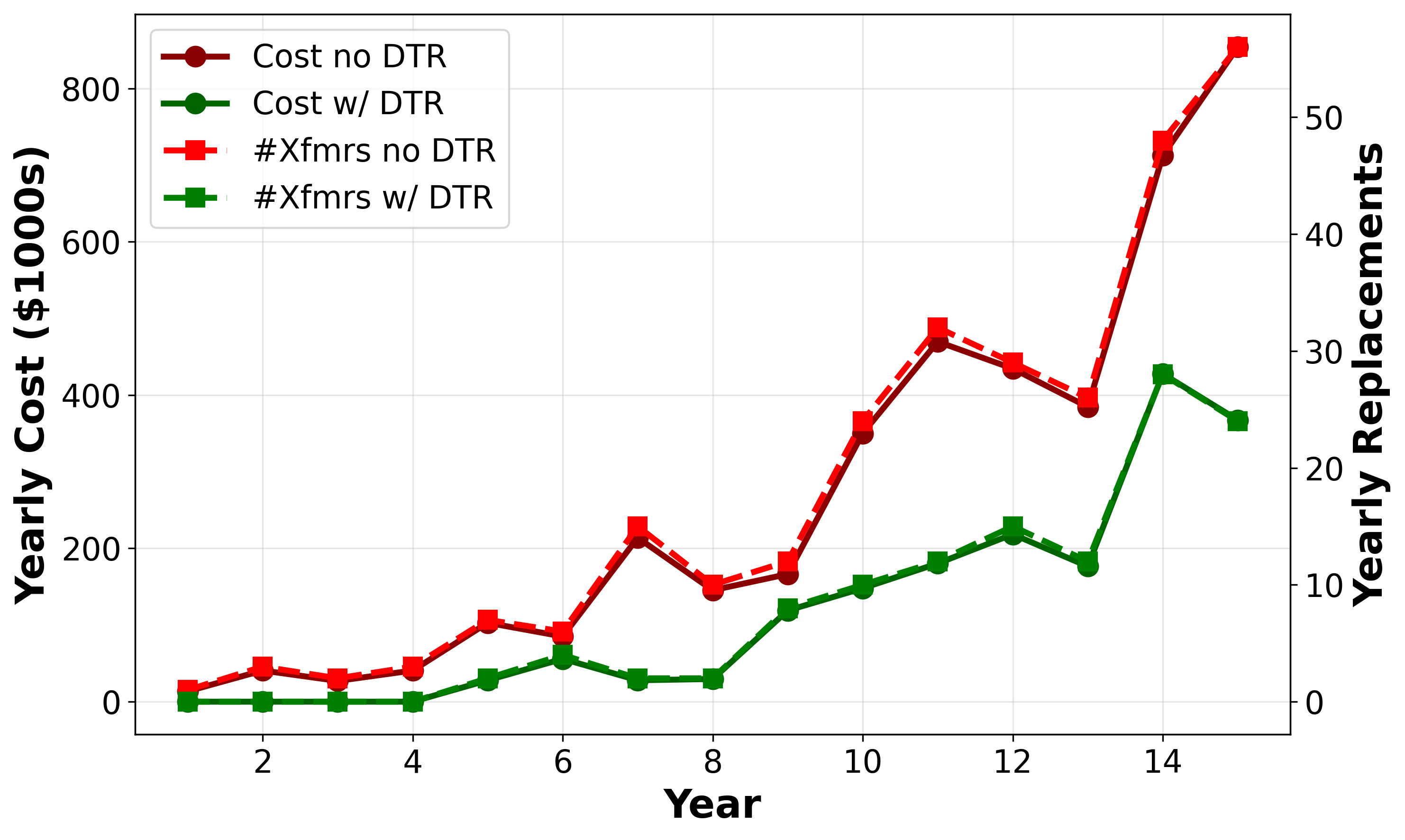}
    \caption{Transformer overloading trend under 5\% load growth without replacement}
    \label{fig:replacement_comparison}
\end{figure}

\begin{figure}[t!]
    \centering
    \includegraphics[scale=0.25]{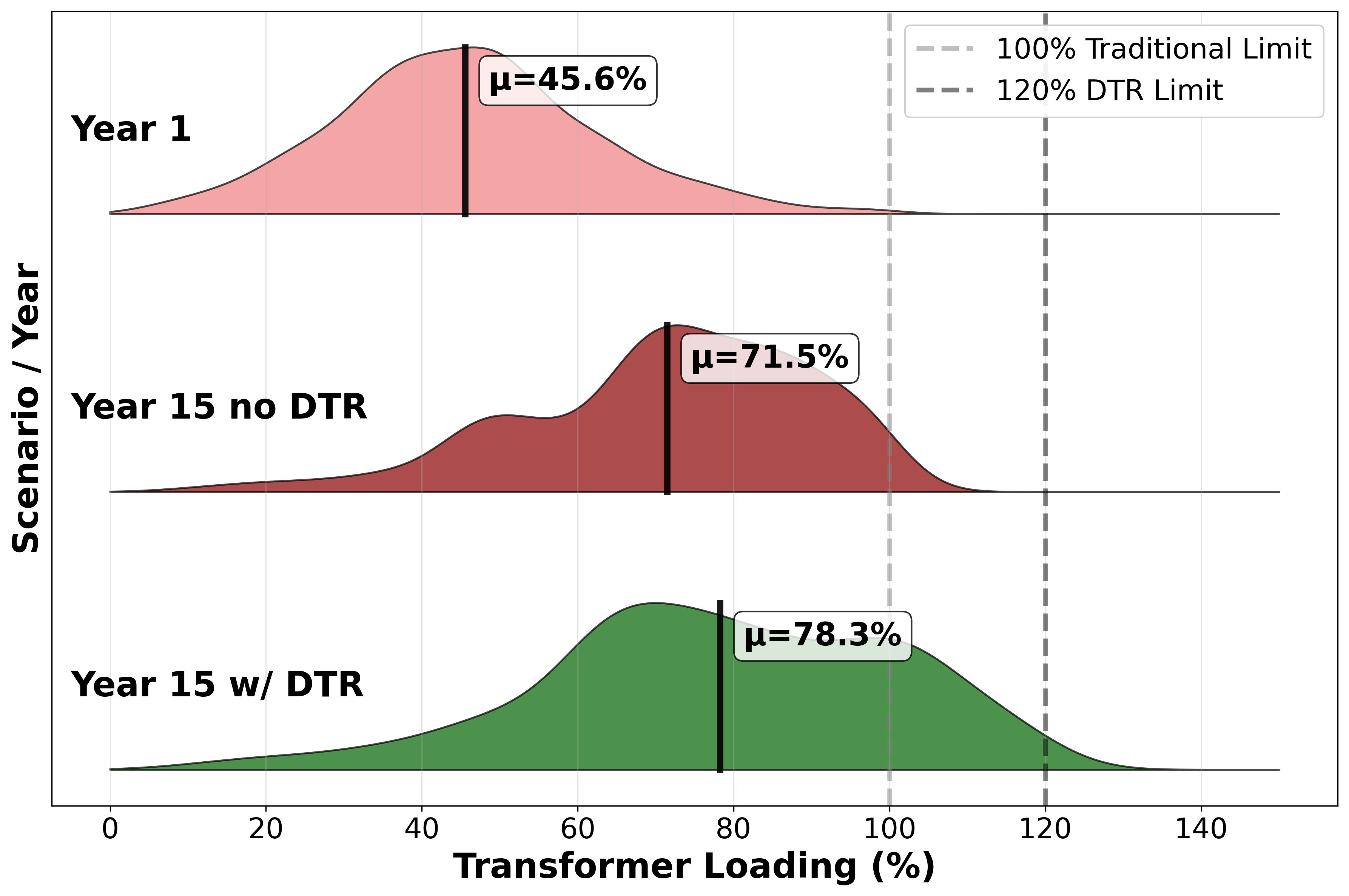}
    \caption{Transformer overloading trend under 5\% load growth without replacement}
    \label{fig:loading_evolution_ridge}
\end{figure}



\subsection{Comparison}


For the DTR case, we apply ambient-adjusted ratings (AAR) to reflect the additional loading capacity available under realistic weather conditions. Specifically, transformer ratings are increased to 120\% of nameplate capacity, consistent with practices in northern states where winter ratings often exceed summer ratings by more than 20\%. A transformer is only replaced when its peak load exceeds the AAR-adjusted rating, rather than the static nameplate rating. When a replacement is triggered, the same sizing rule as in the base case is applied: the smallest standard transformer is selected such that its projected loading remains below 75\% of its AAR-adjusted capacity.

Fig. \ref{fig:replacement_comparison} compares annual costs and replacement counts under traditional upgrade and DTR scenarios. While costs increase steadily over time, within the simulated 15-year horizon (assuming 15 years for the service life of normal DTR monitoring devices), 155 replacements are avoided and approximately \$2.3 million is saved, representing a 56\% cost reduction relative to the traditional 100\% upgrade threshold. On average, annual savings amount to nearly half of the required investment. While the absolute value of savings depends on loading conditions and the total transformer population, the benefits scale with fleet size, making DTR particularly attractive for utilities with large numbers of service transformers.

One other consideration is the accompanying increase in loading levels, as illustrated in Fig. \ref{fig:loading_evolution_ridge}, which shows the projected growth over the same 15-year period, with higher levels observed in the DTR case.

\subsection{Sensitivity With DTR Potential and Growth Rate}

\begin{figure}[t!]
    \centering
    \includegraphics[scale=0.35]{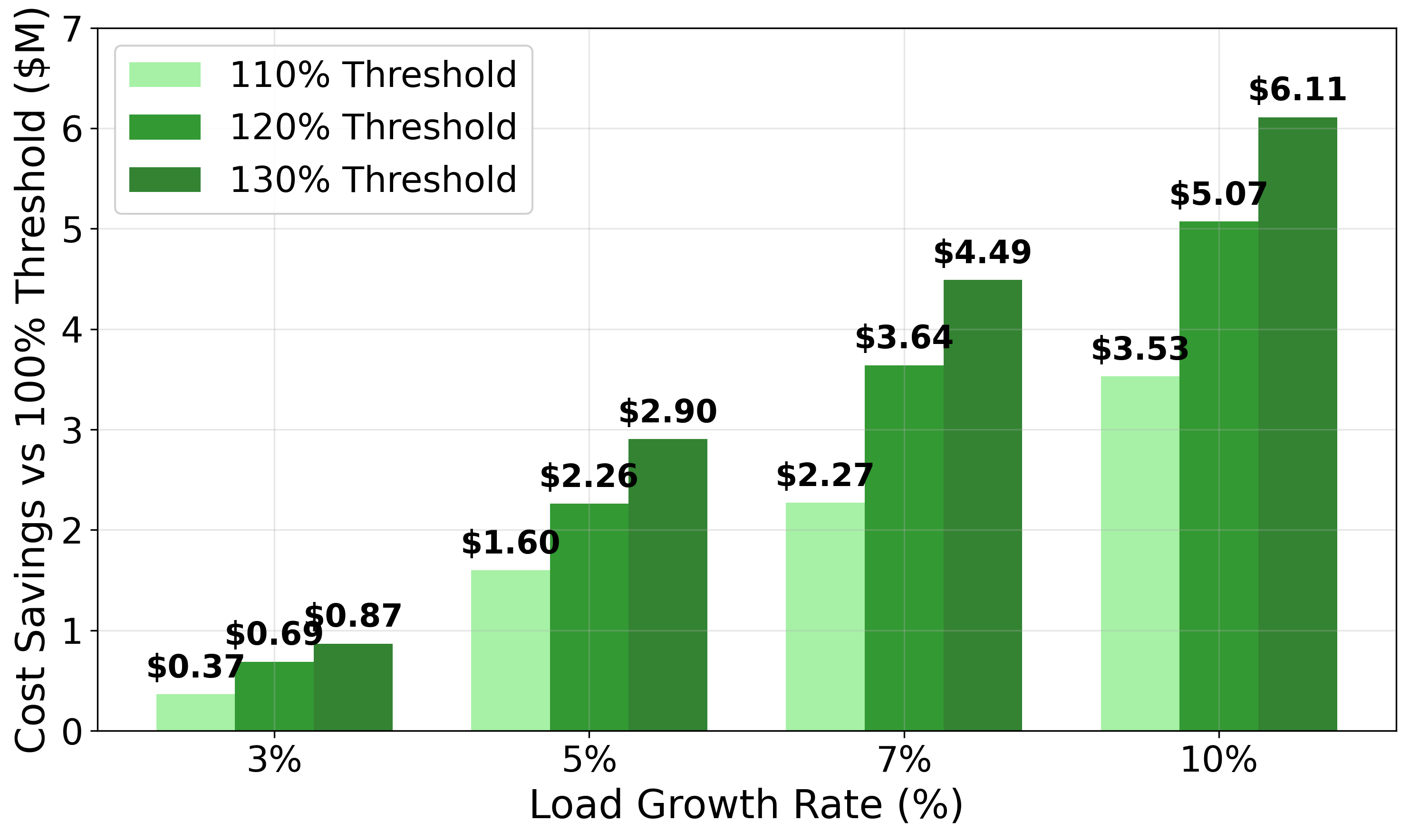}
    \caption{DTR benefits with different load growth sensitivity and thresholds}
    \label{fig:dtr_savings_sensitivity}
\end{figure}

A sensitivity analysis was conducted with thresholds of 110\%, 120\%, and 130\%, and the benefits were compared against the 100\% base case. The resulting savings are shown in Fig. \ref{fig:dtr_savings_sensitivity}. With DTR, utilities can optimize investment strategies and prioritize deployment in regions with greater potential and higher load growth, where the benefits are most significant.

\subsection{Situations Favoring DTR Deployment: When DTR Makes Sense}

Our case study highlights the conditions where DTR offers the most value:  

\begin{itemize}
    \item Pronounced weather diversity, with seasonal or daily temperature variations that can be leveraged.  
    \item Peak demand aligning with cooler periods, allowing higher transformer capacity.  
    \item Sustained or rapid load growth that would otherwise trigger costly upgrades.  
    \item Asset condition and health remaining within acceptable operating limits.  
\end{itemize}

These findings suggest that utilities with winter-peaking systems, accelerated electrification, or diverse climate conditions may be especially well positioned to benefit from DTR adoption.


\section{Discussion}

The practical deployment of DTR or AAR is not without costs. Lessons from dynamic line rating (DLR) in transmission systems show that implementation often requires investments in temperature sensors, communications, data management, and integration with utility operations. For example, one study estimated that equipping a 345-kV line with DLR at three monitoring locations cost approximately \$500,000 \cite{DLR_cost_AEP}. By contrast, applying seasonal AAR to distribution transformers may be far less expensive, potentially requiring only software and database upgrades while leveraging existing weather data. A detailed cost–benefit analysis of AAR/DTR implementation for distribution transformers remains a direction for future work.

In addition, while this analysis focuses on distribution transformers, DTR of overhead lines and cables will also become increasingly important. As transformer capacities are upgraded, system constraints may shift toward lines and cables, highlighting the need for accurate DTR models. For example, \cite{wenbo_dtr_cable} developed a detailed thermal model for underground cables considering unbalanced distribution cables, and a much-needed IEEE guide for the dynamic rating of underground cable systems is currently under development \cite{cable_dtr_std}.

\section{Conclusion}

This paper quantifies the potential benefits of applying DTR technologies in distribution grids, particularly in addressing the accelerated load growth. By using DTR as a short-term non-wires alternative, utilities can boost grid loading capability and defer or avoid costly transformer upgrades. Using a feeder model from the SMART-DS database, our case study showed that in high-potential regions DTR can reduce distribution transformer upgrade costs by nearly half, providing a clear justification for utility DTR programs. Future work will focus on detailed DTR cost assessments and addressing implementation challenges in distribution systems such as interoperability with existing utility programs.




{
\small
\bibliography{citation}

@online{holycross2023,
    author = {{Holy Cross Energy}},
  title        = {2023 Power Supply Roadmap},
  url          = {https://holycross.com/wp-content/uploads/2024/05/HCE_PowerSupplyRoadmap2023_upload.pdf},
    institution = {{Holy Cross Energy}}
}

@techreport{dlr_congress,
  author       = {},
  title        = {Dynamic Line Rating Report to Congress},
  year         = {2019},
  institution  = {U.S. Department of Energy, Washington, D.C., US},
  month        = {June},
  url          = {https://www.energy.gov/oe/articles/dynamic-line-rating-report-congress-june-2019},
  note         = {Accessed: 2024-08-05}
}

@online{DLR_cost_AEP,
  author       = {Shaun Murphy},
  title        = {Simulating the Economic Impact of a Dynamic Line Rating System in {ERCOT}},
  year         = {2018},
  institution  = {CIGRÉ US National Committee},
  url          = {https://cigre-usnc.org/wp-content/uploads/2018/11/04-Simulating-the-Economic-Impact-of-a-Dynamic-Line.pdf},
}

@article{IET_HC_EV_charging,
author = {Zakaria, As’ad and Duan, Chengyan and Djokic, Sasa Z.},
title = {Hosting capacity of distribution networks for controlled and uncontrolled residential EV charging with static and dynamic thermal ratings of network components},
journal = {IET Generation, Transmission \& Distribution},
volume = {18},
number = {6},
pages = {1283-1301},
doi = {https://doi.org/10.1049/gtd2.13025},
year = {2024}
}

@ARTICLE{EVHC_Microgrids,
  author={Lamedica, Regina and Geri, Alberto and Gatta, Fabio Massimo and Sangiovanni, Silvia and Maccioni, Marco and Ruvio, Alessandro},
  journal={IEEE Transactions on Industry Applications}, 
  title={Integrating Electric Vehicles in Microgrids: Overview on Hosting Capacity and New Controls}, 
  year={2019},
  volume={55},
  number={6},
  pages={7338-7346},
  doi={10.1109/TIA.2019.2933800}}

@ARTICLE{EV_impact_2003,
  author={Gomez, J.C. and Morcos, M.M.},
  journal={IEEE Transactions on Power Delivery}, 
  title={Impact of EV battery chargers on the power quality of distribution systems}, 
  year={2003},
  volume={18},
  number={3},
  pages={975-981},
  doi={10.1109/TPWRD.2003.813873}}

@INPROCEEDINGS{EV_impact_limiting_factor,
  author={Sexauer, Jason M. and McBee, Kerry D. and Bloch, Kelly A.},
  booktitle={2011 IEEE Electrical Power and Energy Conference}, 
  title={Applications of probability model to analyze the effects of electric vehicle chargers on distribution transformers}, 
  year={2011},
  volume={},
  number={},
  pages={290-295},
  doi={10.1109/EPEC.2011.6070213}}

@ARTICLE{seasonal_rating_canada,
  author={Awadallah, Mohamed A. and Singh, Birendra N. and Venkatesh, Bala},
  journal={Canadian Journal of Electrical and Computer Engineering}, 
  title={Impact of {EV} Charger Load on Distribution Network Capacity: A Case Study in {Toronto}}, 
  year={2016},
  volume={39},
  number={4},
  pages={268-273},
  doi={10.1109/CJECE.2016.2545925}}

@techreport{paloalto2018,
  title     = {Assessment of CPAU’s Distribution System to Integrate Distributed Energy Resources},
  author    = {{City of Palo Alto Utilities}},
  year      = {2018},
  month     = {April},
  note      = {Memorandum to Utilities Advisory Commission},
}

@techreport{eversource_planning_guide,
  title     = {Eversource Distribution System Planning Guide},
  author    = {{Eversource Energy}},
  year      = {2021},
  url       = {https://www.mass.gov/doc/eversource-distribution-planning-guide/download}
}

@techreport{aep2024,
  title     = {{DER} Technical Interconnection and Interoperability Requirements (TIIR) for the {AEP} System},
  author    = {{American Electric Power Company, Inc.}},
  year      = {2024},
  month     = {February},
  version   = {0.3},
  url       = {https://www.aep.com/assets/docs/requiredpostings/DERInterconnectionRequirements_UpdatedRev3_final_02-01-2024.pdf},
  note      = {Effective Date: 02/01/2024},
}

@ARTICLE{ieee_std_5791,
  author={},
  journal={IEEE Std C57.91-2011 (Revision of IEEE Std C57.91-1995)}, 
  title={{IEEE} Guide for Loading Mineral-Oil-Immersed Transformers and Step-Voltage Regulators}, 
  year={2012},
  volume={},
  number={},
  pages={1-123},
  doi={10.1109/IEEESTD.2012.6166928}}

@ARTICLE{ieee_std_5796,
  author={},
  journal={IEEE Std C57.96-2013 (Revision of IEEE Std C57.96-1999)}, 
  title={{IEEE} Guide for Loading Dry-Type Distribution and Power Transformers}, 
  year={2014},
  volume={},
  number={},
  pages={1-46},
  doi={10.1109/IEEESTD.2014.6725564}}

@article{NREL_supercc,
	title = {High-resolution meteorology with climate change impacts from global climate model data using generative machine learning},
	volume = {9},
	issn = {2058-7546},
	doi = {10.1038/s41560-024-01507-9},
	number = {7},
	journal = {Nature Energy},
	author = {Buster, Grant and Benton, Brandon N. and Glaws, Andrew and King, Ryan N.},
	month = jul,
	year = {2024},
	pages = {894--906},
}

@ARTICLE{wenbo_dtr_cable,
  author={Borbuev, Akim and Wang, Wenbo and Lu, Haowei and Jazebi, Saeed and de León, Francisco},
  journal={IEEE Transactions on Power Delivery}, 
  title={Investment Deferral of Feeder Upgrades Revealed by System-Wide Unbalanced Dynamic Rating: Harvesting the Hidden Capacity of Distribution Systems Discovered by Thermal Map Technology}, 
  year={2021},
  volume={36},
  number={3},
  pages={1594-1602},
  doi={10.1109/TPWRD.2020.3011618}}

@misc{NREL_unit_cost_database,
  author       = {Horowitz, Kelsey},
  title        = {2019 Distribution System Upgrade Unit Cost Database Current Version},
  year         = {2019},
  institution  = {National Renewable Energy Laboratory (NREL)},
  address      = {Golden, CO},
  note         = {Last updated: December 18, 2024},
  doi          = {10.7799/1491263},
  url          = {https://data.nrel.gov/submissions/101}
}

@article{cable_dtr_std,
  title     = {{IEEE Guide for Dynamic Rating of Underground Cable Systems (IEEE Std 3174)}},
  author    = {},
  journal   = {IEEE Standards Association},
  year      = {2023},           
  url       = {https://standards.ieee.org/ieee/3174/11278/},
  note      = {Accessed: April 10, 2025}
}

@online{eversource,
  author       = {{Eversource Energy}},
  title        = {Distributed Energy Resources Project Costs},
  year         = {2022},
  url          = {https://www.eversource.com/content/residential/about/doing-business-with-us/interconnections/massachusetts/distributed-energy-resources-project-costs},
  note         = {Accessed: February 10, 2025},
}

@online{sce,
  author       = {{Southern California Edison (SCE)}},
  title        = {Unit Cost Guide},
  year         = {2022},
  url          = {https://www.sce.com/sites/default/files/custom-files/Web%20files/Attachment_A-Unit_Cost_Guide.pdf},
  note         = {Accessed: February 10, 2025},
}

@techreport{Mai_NREL_EFS,
  author      = {Trieu Mai and Paige Jadun and Jeffrey Logan and Colin McMillan and Matteo Muratori and Daniel Steinberg and Laura Vimmerstedt and Ryan Jones and Benjamin Haley and Brent Nelson},
  title       = {Electrification Futures Study: Scenarios of Electric Technology Adoption and Power Consumption for the United States},
  institution = {National Renewable Energy Laboratory (NREL)},
  address     = {Golden, CO, USA},
  year        = {2018},
  number      = {NREL/TP-6A20-71500},
  url         = {https://www.nrel.gov/docs/fy18osti/71500.pdf},
}
\bibliographystyle{IEEEtranN}
}
\end{document}